\documentclass[10pt]{llncs}
\usepackage{cite}
\usepackage{amsmath,amssymb,amsfonts}
\usepackage{algorithmic}
\usepackage{graphicx}
\usepackage{textcomp}
\usepackage{url}
\usepackage{xcolor}
\usepackage{todonotes}
\usepackage[hidelinks]{hyperref}
\usepackage{cleveref}

\begin{document}

\newcommand{\qkdvendora}{ThinkQuantum} 
\newcommand{\qkdvendorb}{Quantum Optics Jena} 
\newcommand{\citya}{Nordhausen} 
\newcommand{\cityb}{Sundhausen} 

\title{PQC-Enhanced QKD Networks: A Layered Approach\thanks{This is the full version of a paper which appears in the IEEE International Conference on Quantum Communications, Networking, and Computing (QCNC), Kobe, Japan, April 6--8, 2026. \textcopyright~IEEE, 2026.}}
\author{
  Paul Spooren\inst{1}
\and
Andreas Neuhold\inst{2}
\and
Sebastian Ramacher\inst{3}
\and
Thomas H\"uhn\inst{1}
}

\institute{
  University of Applied Sciences Nordhausen, Nordhausen, Germany \email{firstname.lastname@hs-nordhausen.de} \and
  CANCOM Converged Services GmbH, Graz, Austria \email{Andreas.Neuhold@cancom.com} \and
  AIT Austrian Institute of Technolgy, Vienna, Austria \email{sebastian.ramacher@ait.ac.at}
}

\maketitle

\begin{abstract}
We present a layered and modular network architecture that combines Quantum Key Distribution (QKD) and Post-Quantum Cryptography (PQC) to provide scalable end-to-end security across long distance multi-hop, trusted-node quantum networks.
To ensure interoperability and efficient practical deployment, hop-wise tunnels between physically secured nodes are protected by WireGuard with periodically rotated pre-shared keys sourced via the ETSI GS QKD 014 interface. On top, Rosenpass performs a PQC key exchange to establish an end-to-end data channel without modifying deployed QKD devices or network protocols. This dual-layer composition yields post-quantum forward secrecy and authenticity under practical assumptions. We implement the design using open-source components and validate and evaluate it in simulated and lab test-beds. Experiments show uninterrupted operation over multi-hop paths, low resource footprint and fail-safe mechanisms. We further discuss the design's compositional security, wherein the security of each individual component is preserved under their combination and outline migration paths for operators integrating QKD-aware overlays in existing infrastructures.
\end{abstract}


\section{Introduction}
\label{sec:introduction}

Quantum computing development threatens the foundations of current public-key cryptography. Algorithms such as RSA, DSA, and elliptic-curve cryptography depend on mathematical problems that become breakable under Shor's algorithm on a sufficiently large quantum computer~\cite{DBLP:conf/focs/Shor94}. This renders conventional pre-quantum asymmetric cryptography insecure in a post-quantum context.

Post-Quantum Cryptography (PQC) is already being integrated into production systems: the Signal Messenger protocol now supports hybrid key exchanges using quantum-resistant primitives~\cite{signal_pqc}; Apple's iMessage introduced PQC-based encryption~\cite{apple_pq3}; OpenSSH has incorporated ML-KEM into its handshake process~\cite{openssh99}; and drafts for hybrid PQC-TLS are under active discussion at the IETF while already deployed by Google, Cloudflare, and others~\cite{cloudflare}. These developments demonstrate the practical viability and interoperability of PQC with existing network protocols.

As an alternative to this approach, Quantum Key Distribution (QKD) offers a fundamentally different approach to secure communication, rooted not in mathematical hardness assumptions but in the laws of quantum physics~\cite{alleaume2014using}. First proposed by Bennett and Brassard in 1984~\cite{DBLP:conf/crypto/BennettB84}, QKD enables the generation of shared symmetric keys between two parties by transmitting quantum states, such that any eavesdropping attempt introduces detectable disturbances. This enables the construction of cryptographic systems offering information-theoretic secrecy~\cite{renner2023quantum}. To make such systems scalable beyond point-to-point links,  an additional layer of Key Management Systems (KMSs) on top of internal QKD KMSs, orchestrated by Software Defined Networks (SDNs), is required to relay key material across intermediate Trusted Nodes (TNs), allowing end nodes to access and use the keys within higher-layer applications for any pair of nodes in the network~\cite{mehic2020,huttner2022long,DBLP:conf/icton/MartinBOBVSSACSSEDRPL23}. Trusted Nodes (TNs) are intermediate nodes that must be physically and operationally hardened to protect keys from unauthorized access; unlike ideal quantum repeaters, they introduce a trust assumption.

\subsection{Problem Statement}
\label{sec:problem-statement}

While PQC offers software-based, end-to-end protection resistant to quantum attacks and deployable on existing infrastructures, its security depends on evolving algorithmic standards and hardness assumptions believed to be secure against quantum computers~\cite{castryck2023efficient}.
QKD, on the other hand, provides information-theoretic keys but faces intrinsic scalability and operational constraints~\cite{anssi2024} yet multi-hop deployments rely on trusted nodes and interoperable key delivery. Each TN becomes a potential point of exposure, nullifying end-to-end secrecy and demanding extensive physical protection.
A KMS may internally employ PQC mechanisms to strengthen the synchronization of key material; however, the current interoperability standards for KMSs do not specify PQC support~\cite{ETSI020}. Consequently, security gaps can arise at the interfaces between heterogeneous KMS implementations, for instance, at border nodes or between institutions that interconnect their data links.
Beyond the security aspects, deploying QKD networks is complicated by the need of operating a separate KMS layer and SDN infrastructure whereas some of the standards for interoperability are still in development and systems are not widely available.

\subsection{Contributions}
\label{sec:contributions}

This work introduces and evaluates a modular \emph{KMS-free} architecture that unifies classical, post-quantum, and quantum cryptography into a single tri-layer framework. With KMS-free we refer to the removal of the additional network-level Key Management System layer used for forwarding keys and interoperability between QKD and the need for an orchestrating SDN providing networks paths to the extra KMS layer. It does not remove the local KMS components built into individual QKD devices, which remain necessary to provide key material via standardized interfaces such as ETSI GS QKD 014 on single links. By combining these paradigms, the design achieves QKD and PQC based forward-secure communication across multi-hop, quantum-aware networks. The main contributions are:
\begin{enumerate}
  \item \textbf{Composable layering:} Design and implementation of a KMS-free network framework that combines end-to-end PQC encryption across hop-wise tunnels secured by QKD-derived PSKs. The architecture maintains clean separation between link-layer and end-to-end security functions. All operations comply with the ETSI GS QKD 014~\cite{ETSI014} standard to access QKD keys without the need to adapt any of the standardized interfaces.
  \item \textbf{Layer Separation:} Forward secrecy is achieved by independent key rotation at both the QKD hop layer and the PQC end-to-end layer. This separation limits the impact of any single compromise and protects against active and passive attacks, specifically, scenarios in which an attacker exploits vulnerabilities in post-quantum or classical primitives.
  \item \textbf{Implementation and Experimental Evaluation:} Open-source prototype and lab evaluation with resource, stability and scalability measurements.
  \item \textbf{Operational guidance:} Migration path for operators; discussion of assumptions, limits, and failure modes.
\end{enumerate}

\subsection{Related Work}
\label{sec:related_work}

The integration of Quantum Key Distribution (QKD) with Post-Quantum Cryptography (PQC) has been explored in several prior studies. This work extends that line of research by presenting a modular and KMS-free architecture that combines the two domains without modifying existing QKD or network protocols.

The open-source project Arnika~\cite{githubGitHubArnikaprojectarnika} presents an option to integrate QKD and PQC keys into a Wireguard based VPN. It utilizes the ETSI GS QKD 014 API and the PQC software Rosenpass~\cite{rosenpass_whitepaper} to derive hybrid keys through a key-derivation function. However, Arnika depends on an extra  Key Management System (KMS) layer  for key forwarding.

Other research focuses on hybrid key exchange mechanisms. The Muckle~\cite{DBLP:conf/pqcrypto/DowlingHP20}, Muckle+~\cite{DBLP:conf/pqcrypto/BrucknerRS23}, and Muckle\#~\cite{battarbee2024quantumsafehybridkeyexchanges} protocols merge QKD-derived entropy with classical and post-quantum primitives to build authenticated key exchanges. These efforts lay the groundwork for future PQC/QKD-hybrid protocols such as PQC-QKD-TLS, but they require protocol-level modifications as also highlighted in VMuckle for MACsec~\cite{buruaga2025vmuckle}.

Chen et al.~\cite{DBLP:conf/iccnc/ChenXLLY24} propose integrating QKD keys directly into the WireGuard protocol by replacing its KEM with a QKD-enhanced variant that combines classical and quantum keys using HMAC evaluation. While effective, this approach requires modification of all deployed WireGuard implementations, complicating adoption in existing infrastructures. In contrast, Hülsing et al.~\cite{DBLP:conf/sp/HulsingNSWZ21} introduce a PQC-based key exchange designed to replace WireGuard’s KEM entirely. A practical implementation and progression of that concept is Rosenpass, which realizes the proposed mechanism externally and injects PQC-derived keys as pre-shared keys into unmodified WireGuard instances.

Further work has investigated network-level management of QKD infrastructures. Mehic et al.~\cite[p. 4]{mehic2020} and Huttner et al.~\cite{huttner2022long} demonstrate software-defined network (SDN) control frameworks for large-scale QKD systems. James et al.~\cite{DBLP:conf/IEEEares/JamesLRT23} analyze KMS designs for multi-hop QKD networks and describe daisy-chain-style key forwarding schemes—both centralized and distributed—using XOR-based aggregation. Building on this, Franzoi et al.~\cite{franzoi2025secure} formalize key forwarding within a cryptographic model that defines security properties for QKD networks with TNs.

Sáez et al.~\cite{saez2024current} perform a comprehensive gap analysis for Quantum Key Distribution Networks (QKDNs), highlighting interoperability challenges and emphasizing the importance of standardized interfaces and PQC integration for future network deployments.

The approach presented in this paper departs from these prior designs by introducing a modular, KMS-free layering model: QKD secures individual hops via ETSI-014 interfaces, while PQC provides end-to-end confidentiality through Rosenpass. This architecture achieves pragmatic forward secrecy within existing, distance-limited QKD networks and aligns with standardized telecom-grade interfaces already familiar to network operators~\cite{clark1988design}.

\section{System and Threat Model}
\label{sec:system-model}

The considered QKD network consists of multiple physically secured trusted nodes (TNs), connected via pairwise QKD links. Each TN maintains authenticated key exchanges in compliance with ETSI-014, enabling retrieval of symmetric key material for directly connected neighbors. QKD devices are assumed secure and to correctly implement ETSI-014 authentication and integrity mechanisms, ensuring that key establishment and signaling remain protected against manipulation. All TNs are deployed in physically secured facilities with controlled access and continuous monitoring. End nodes are geographically separated and cannot directly exchange QKD-derived key material; instead, they are interconnected through at least one intermediate TN.

We consider multiple adversarial models, each targeting different layers of the system. A passive quantum-capable adversary can observe all classical and quantum channels, perform eavesdropping, and is assumed capable of breaking classical and PQC asymmetric cryptography but not compromising trusted nodes; QKD provides protection in this case. An adversary may compromise individual trusted nodes and access their local key material but cannot break PQC primitives; Rosenpass provides protection under this model. Finally, an adversary capable of compromising both trusted nodes and PQC primitives is considered, but this adversary lacks quantum advantage and therefore cannot break the classical cryptography used by WireGuard.

The attacker's objective is to break ciphertext indistinguishability, i.e., to distinguish between encryption of different messages and thereby recover information about end-to-end session keys or exchanged data. We do not consider adversaries causing prolonged disruption of communication, e.g., via Distributed Denial of Service attacks. Furthermore, our objective is not to achieve information-theoretic end-to-end security, which would require quantum repeater functionality that lacks current technological readiness~\cite{krutyanskiy2023telecom}.

\section{Design}
\label{sec:design}

The architecture follows a strict layering principle that combines classical, quantum, and post-quantum cryptography while maintaining clear separation of their functions. At the lower layer, WireGuard provides classical encryption, with its pre-shared keys periodically refreshed using QKD-derived key material to secure hop-wise tunnels between physically guarded nodes. Above this, an end-to-end PQC handshakes through the QKD-protected backbone generates key material, which secures an end-to-end WireGuard tunnel for application data. This layered composition preserves composability and confines potential failures within each cryptographic domain.

\subsection{Layering Principle}

\begin{figure}
    \centering
    \includegraphics[width=1\linewidth]{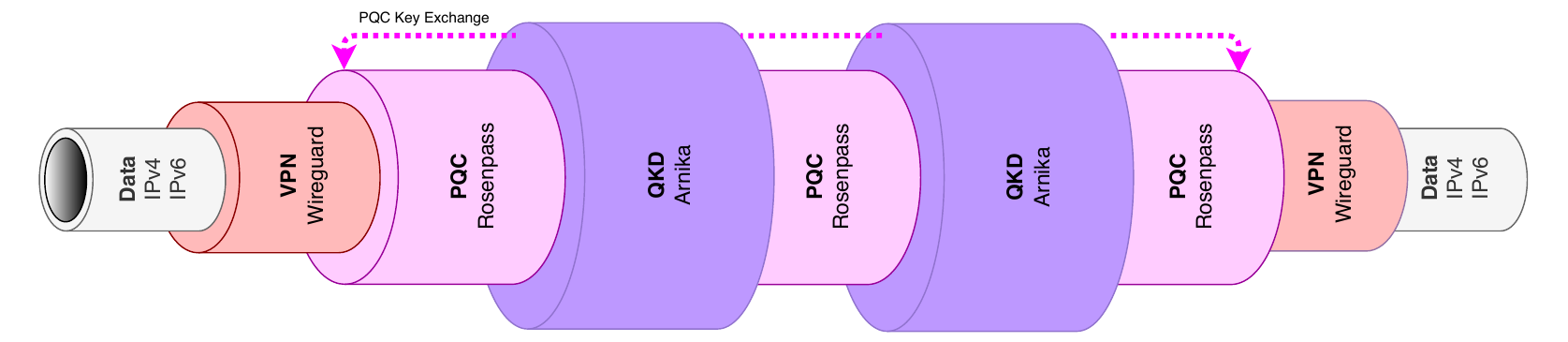}
    \caption{Conceptual schema of the cryptographic layers}
    \label{fig:daisypass-pipes}
\end{figure}

\Cref{fig:daisypass-pipes} conceptually illustrates the design layers. At the hop layer (purple), each link between adjacent TNs is implemented as a WireGuard tunnel. The pre-shared key (PSK) for these tunnels is provided using ETSI GS QKD 014 interfaces from local QKD modules. This QKD-secured WireGuard link ensures confidentiality and authentication between directly connected nodes without  requiring an extra KMS layer with key-forwarding capabilities or an SDN to provide network paths.

At the end-to-end layer (pink), a Rosenpass-based PQC handshake is conducted over the concatenated chain of QKD-secured WireGuard tunnels. The generated PQC-based key material secures the end-to-end overlay (dark red), which is usable for application data (grey). Intermediate TNs provide transport services only.

\subsection{Routing and Composition}

\begin{figure}
    \centering
    \includegraphics[width=1\linewidth]{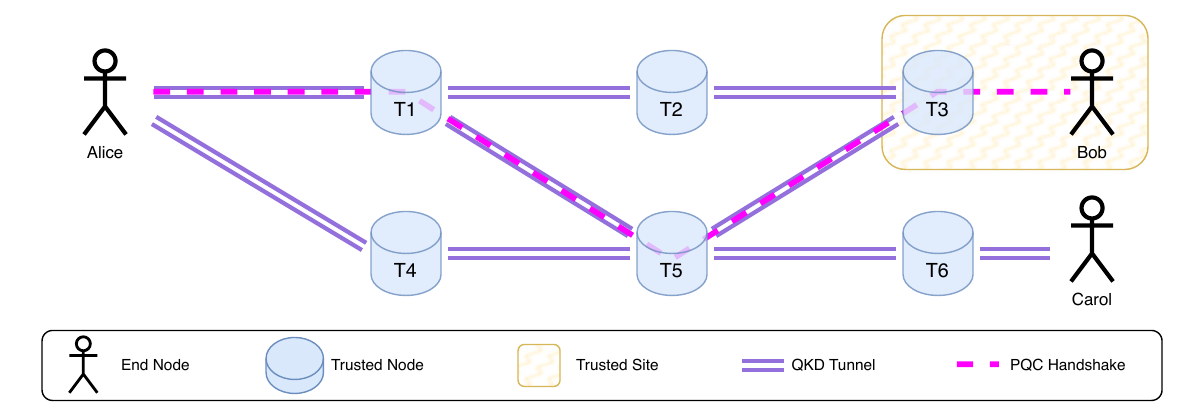}
    \caption{Schematic representation of a layered network}
    \label{fig:network}
\end{figure}

\Cref{fig:network} illustrates the schematic structure of the layered design described above. The end nodes Alice, Bob, and Carol are separated by distances that exceed direct QKD connectivity. They therefore communicate through intermediate QKD-secured tunnels connecting multiple TNs. In this configuration, Alice and Bob perform a PQC key exchange over the QKD-protected path $\mathsf{T1} \rightarrow \mathsf{T5} \rightarrow \mathsf{T3}$, then establishing a secure data channel (not explicitly shown).

The figure also illustrates a \textit{last-mile} scenario in which Bob does not operate a dedicated QKD device. Instead, Bob connects to the QKD-secured backbone through the trusted node $\mathsf{T3}$ located within a trusted site. Such a site could be a factory, hospital, or embassy that deploys a single QKD device to secure all external communications.

Additionally, Alice and Bob can simultaneously initiate PQC key exchanges with Carol. Because the QKD-secured tunnels carry standard network traffic, they inherently support parallel connections, enabling multiple independent PQC exchanges to coexist over the same infrastructure. Due to multi-tenancy, security could be further strengthened through a multi-link transmission approach for PQC data traffic. Alice and Bob can establish multiple independent PQC handshakes across diverse paths (e.g., $\mathsf{T1} \rightarrow \mathsf{T2} \rightarrow \mathsf{T3}$ and $\mathsf{T4} \rightarrow \mathsf{T5} \rightarrow \mathsf{T3} \rightarrow \mathsf{T4}$). This multi-link approach would force an adversary to compromise two trusted nodes.

Failure of individual TNs or QKD tunnels does not necessarily cause a disruption of the PQC handshakes or data exchange. By employing conventional routing mechanisms, the network could dynamically reroute through alternative QKD-secured paths. For instance, if node $\mathsf{T1}$ becomes unavailable, node $\mathsf{T4}$ can continue its role.

\section{Implementation}
\label{sec:implementation}

This section provides an overview of the system components employed in the proposed design, along with a detailed explanation of their interactions.

\subsection{Component Overview}

\subsubsection{WireGuard}

In recent years, WireGuard~\cite{donenfeld2017wireguard} has emerged as a major contender among VPN solutions, combining ease of configuration with high performance, surpassing tools such as OpenVPN~\cite{DBLP:conf/codaspy/MackeyMNVC20}. Although the symmetric encryption scheme based on ChaCha20~\cite{bernstein2008chacha} remains resistant to quantum attacks, the key exchange mechanism using Curve25519~\cite{bernstein2006curve25519} does not. To mitigate this weakness, WireGuard allows the inclusion of a pre-shared key (PSK), providing additional protection even if the Curve25519-based key exchange were compromised. However, this approach alone does not ensure forward secrecy, as exposure of the PSK would compromise all previously exchanged data. The design presented in this paper introduces a mechanism that achieves forward secrecy through periodic PSK rotation using QKD and PQC.

\subsubsection{ETSI GS QKD 014 (ETSI-014)}

The deployment of QKD networks is guided by several standards and recommendations defining overall architecture~\cite{recommendation20193800,recommendation3801,recommendation3802,recommendation3803} and the interfaces between software and hardware components~\cite{ETSI004,ETSI014,ETSI015,ETSI016,ETSI020}. For applications utilizing QKD keys, the most relevant specification is ETSI GS QKD 014~\cite{ETSI014}, which defines an API for KMSs and QKD devices, enabling users to retrieve key material. In this model, the initiator obtains a key and its identifier (key ID) from the local KMS, transmits the key ID to the peer, and the recipient then requests the corresponding key from its own local KMS via the same key ID.

\subsubsection{Arnika}

The Arnika utility acts as an intermediary between the QKD hardware (via ETSI-014 APIs) and WireGuard. It retrieves key material from the QKD device and injects it into WireGuard as a pre-shared key. Arinka performs PSK injection at defined intervals to maintain forward secrecy and injects random keys if retrival of QKD keys fails for specified time.

\subsubsection{Rosenpass}

Rosenpass~\cite{rosenpass_whitepaper} represents the practical implementation and evolution of the \textit{Post-Quantum WireGuard} concept introduced by Hülsing et al.~\cite{DBLP:conf/sp/HulsingNSWZ21}. It employs a combination of Classic McEliece~\cite{bernstein2017classic}, a fourth-round candidate in the NIST PQC competition, and CRYSTALS-Kyber~\cite{DBLP:conf/eurosp/BosDKLLSSSS18}, the selected winner in the same process. These two well-reviewed PQC algorithms provide mechanisms for authentication and confidentiality. The keys exchanged through Rosenpass can be injected into a running WireGuard instance as pre-shared keys, protecting sessions against quantum adversaries as long as the underlying PQC assumptions hold. Multiple Rosenpass instances can operate in parallel, allowing independent applications to maintain separate PQC key material for distinct sessions. As with Arnika, consecutive failures to exchange messages and generate a new PSKs, cause Rosenpass to inject a random keys to disrupt WireGuards data exchange.

Although scalability was not an objective of this study, we conducted a simulation in which a Rosenpass instance served 5,000 peers on a single Intel Core i9-13900H core. All peers completed the key exchange within the default 120s interval. Additional details are provided in the supplementary materials.

\subsection{Integration Workflow}

\begin{figure}
    \centering
    \includegraphics[width=1\linewidth]{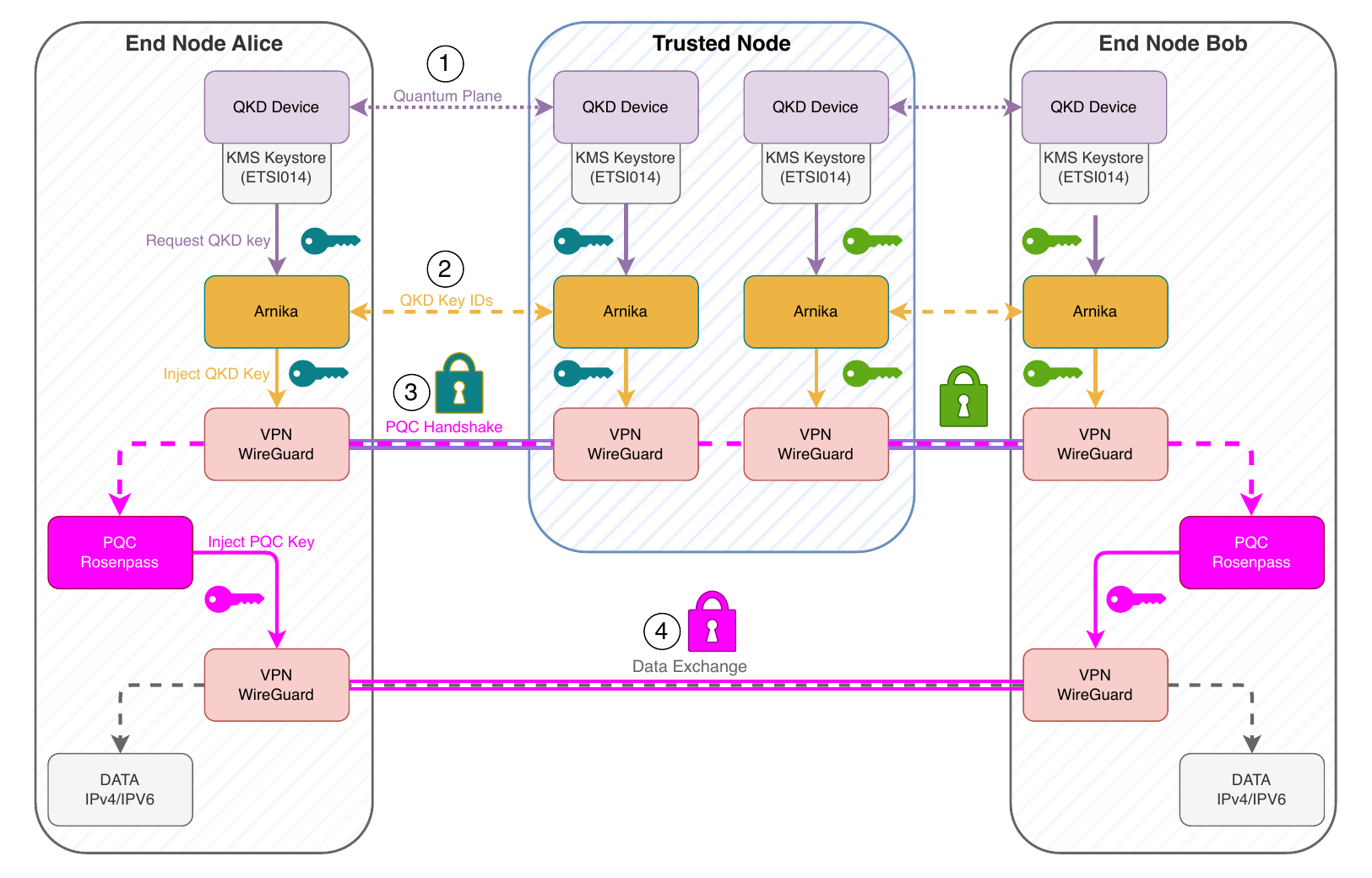}
    \caption{Multi-hop setup with Alice, Bob and a trusted node in-between}
    \label{fig:e2e}
\end{figure}

\Cref{fig:e2e} illustrates all stages of the setup, including the utility Arnika, Rosenpass, and WireGuard. First, the QKD devices generate key material \textbf{(1)} and make it available via the ETSI GS QKD 014 API. The Arnika utility \textbf{(2)} consumes this key material, negotiates the active key ID with neighboring nodes, and injects the corresponding key material as a pre-shared key (PSK) into the hop-level WireGuard tunnel \textbf{(3)}. Rosenpass routes its traffic through the QKD-secured WireGuard tunnels \textbf{(3)} to perform PQC key exchanges. The resulting key material is injected into the final WireGuard data tunnel \textbf{(4)}, which enables Alice and Bob to exchange data securely.

\subsection{Fail-Safe Mechanism}

When the QKD layer becomes unavailable a fail-safe mechanism prevents downgrade attacks, where the data tunnel is interrupted after a defined time. WireGuard, Arnika, and Rosenpass each refresh their PSK or session key every 120s, with a 60s grace window before terminating a connection (WireGuard) or injecting random keys (Arnika, Rosenpass). The components operate independently, so Arnika may inject keys at any point within WireGuard's \textbf{(3)} 120s handshake cycle, and Rosenpass may inject keys at arbitrary points within WireGuard data \textbf{(4)} handshake. Loss of QKD hardware therefore does not trigger an immediate data-path outage; each layer fails according to the time elapsed since its last successful handshake. A component stops no earlier than 60s after the previous layer fails and no later than 180s. With the architecture in \Cref{fig:e2e}, loss of the QKD plane \textbf{(1)} propagates to a full data-path \textbf{(4)} interruption after 240s to 720s.

\section{Evaluation}

\subsection{Simulation and Emulation Results}
\label{sec:experiments}

This section presents six experiments to validate the design, evaluate the performance, scalability, and practicality of the combined QKD–PQC architecture under simulated and lab conditions. The experiments cover system scalability, software efficiency, resource usage and cryptographic agility. Experiment 1 employs an initial Python proof-of-concept called \textit{quPSKd} for QKD-to-WireGuard key injection and only used to validate the concept of daisy-chaining QKD-tunnels. All other experiments use the open source tool \textit{Arnika}, following the paper scope to use existing software.

The experiments 2 to 5 are simulations and ran on a Intel Core i9-13900H machine with Ubuntu 24.04.3 LTS (Kernel 6.8.0-87-generic) and Containerlab 0.71.1. All datasets collected during performance and stability evaluations, including raw measurement logs, key rotation traces, and system resource utilization data, are accessible in the supplementary files to ensure transparency and reproducibility of results\footnote{\url{https://github.com/aparcar/qkd-pqc-paper-supplementary-files}}.

\subsubsection{Test 1 – Prototype Validation}

\begin{figure}
    \centering
    \includegraphics[width=1\linewidth]{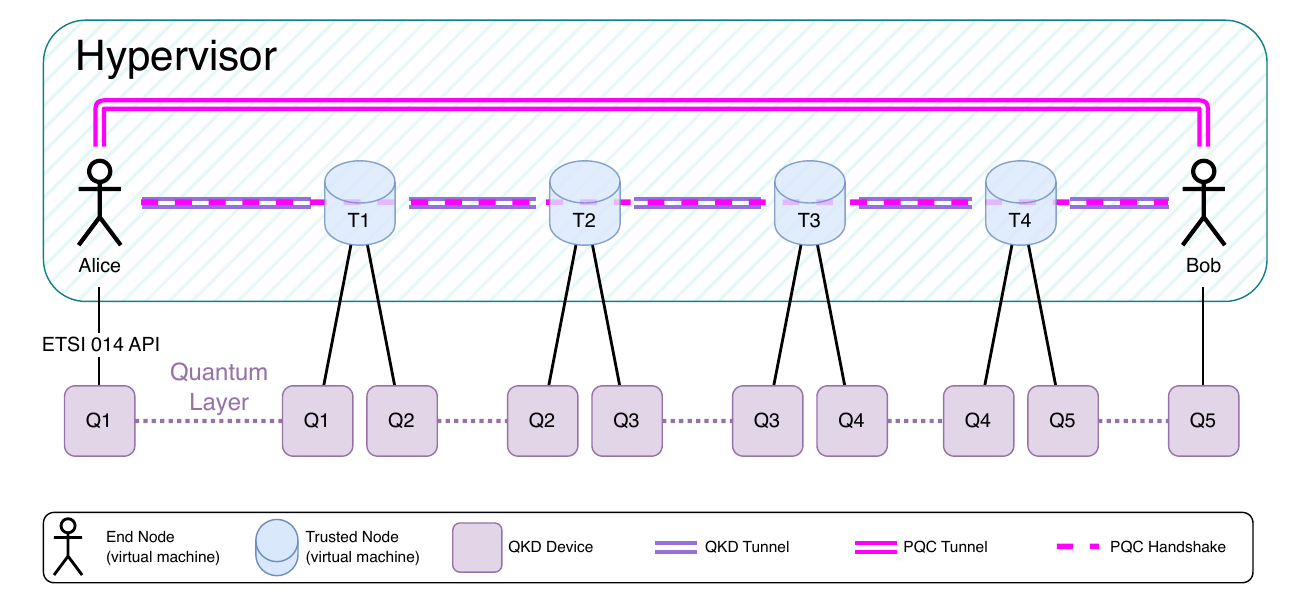}
    \caption{Testbed configuration for evaluating the design}
    \label{fig:testbed}
\end{figure}

The initial prototype validated the feasibility and scalability of the layered QKD–PQC design using physical QKD hardware. The testbed comprised a mixture of ten QKD devices from \qkdvendora~ and \qkdvendorb. Two end nodes were connected via four intermediate TNs in a daisy-chain topology. Each node had access to a QKD device through the ETSI GS QKD 014 API and established WireGuard tunnels secured with QKD-derived keys. Figure~\ref{fig:testbed} shows the experimental setup.

The prototype software \textit{quPSKd}  successfully created five intermediate tunnels for the Rosenpass-based PQC handshakes across all nodes. Automated scripts rebooted and reconnected the entire setup for multiple iterations while capturing timing and data exchange metrics. Results confirmed stable multi-hop key exchanges and demonstrated that the layered tunnel approach scales across multiple physical nodes.

Furthermore the experiment tracked the bandwidth usage of links connecting trusted nodes, including Arnikas key negotiation, WireGuards handshakes as well as the encrypted data over the WireGuard tunnel, the PQC handshake. \Cref{tab:packets_bytes} presents the number of packets and traffic in Bytes.

\begin{table}[!t]
    \caption{Packets and Traffic per Handshake or Key Negotiation}
    \label{tab:packets_bytes}
    \centering
    \begin{tabular}{lccc}
        \hline
            Software & Packet Number & Traffic (bytes) & Type \\
        \hline
            WireGuard & 3 & 398 & UDP \\
            Arnika & 2 & 78 & TCP \\
            Rosenpass & 4 & 4772 & UDP \\
        \hline
    \end{tabular}
\end{table}

\subsubsection{Test 2 – Long Distance}

The second experiment extended the prototype to paths of 10 and 100 trusted nodes. End nodes, trusted nodes, and QKD devices were emulated in Linux containers, with each QKD container implementing the ETSI 014 interface and delivering pre-generated keys. Apart from the increased path length, the topology matched that of the first experiment. The metric of interest was the time required to bring up all QKD tunnels, derive a PQC key over those tunnels, configure the PQC-secured data channel, and achieve successful end-to-end communication.

This simulation assessed the architecture from a networking standpoint. In contrast to the conventional Internet, where fiber repeaters do not increase hop count, long terrestrial QKD links introduce additional hops. To accommodate the extended path length, the Linux kernel hop limit was set to 100,\footnote{\texttt{sysctl -w net.ipv4.ip\_default\_ttl=100}} exceeding the default capacity for 64 intermediate hops.

Across 100 runs, the mean setup time was 10.27s for 10 intermediate nodes and 10.62 s for 100 intermediate nodes, indicating that overall initialization latency is dominated by the slowest QKD hop rather than cumulative hop count.

\subsubsection{Test 3 – Enforced Dual QKD Link \& Cryptographic Agility}

\begin{figure}
    \centering
    \includegraphics[width=1\linewidth]{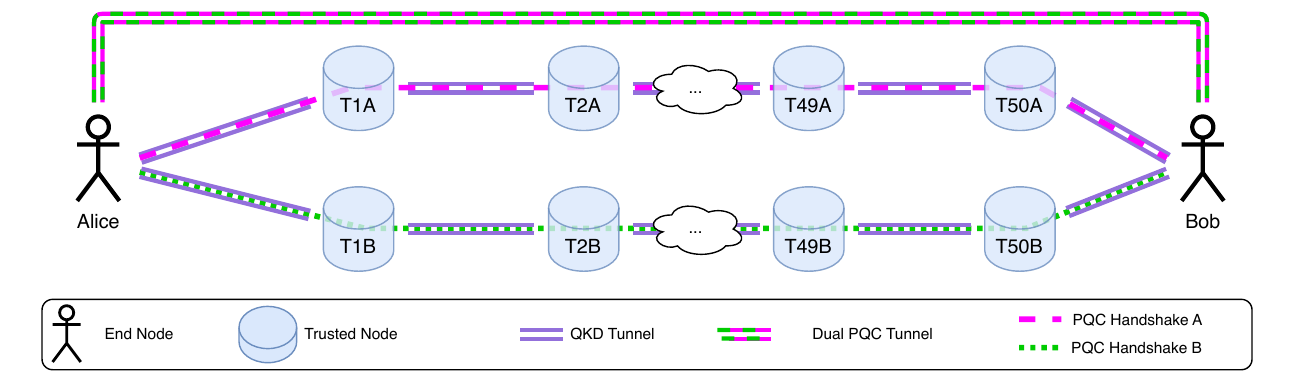}
    \caption{Simulated testbed featuring two QKD daisy chains and two PQC handshake implementations}
    \label{fig:testbed-multipath}
\end{figure}

\Cref{fig:testbed-multipath} presents a modification of the prior simulation. The topology, including the QKD simulators, remained unchanged, but two daisy chains of 50 intermediate nodes were added to supply two independent paths for PQC key exchanges.

Two Rosenpass versions\footnote{Release v0.2.2 and the latest commit at the time of testing, \texttt{582d2735}.} were deployed on each end node. Their implementation differences make them mutually incompatible, so one QKD path was assigned to version 0.2.2 and the other to the development version. Each end node thus derived two independent 32-byte PSKs, which were combined with a hash function, and the resulting key was injected into the WireGuard data tunnel.

The simulation examined the concurrent use of two QKD daisy chains, increasing the adversarial burden from compromising a single trusted node to compromising one in both path. It also demonstrated functional decoupling between the QKD transport layer and the PQC handshake: end-node software can be exchanged without any modifications to intermediate nodes.

Across 100 runs, the mean time to provision a data tunnel secured by two independent QKD paths and two PQC mechanisms was 9.93s.

\subsubsection{Test 4 – Degraded Link Connectivity}

\begin{figure}
    \centering
    \includegraphics[width=1\linewidth]{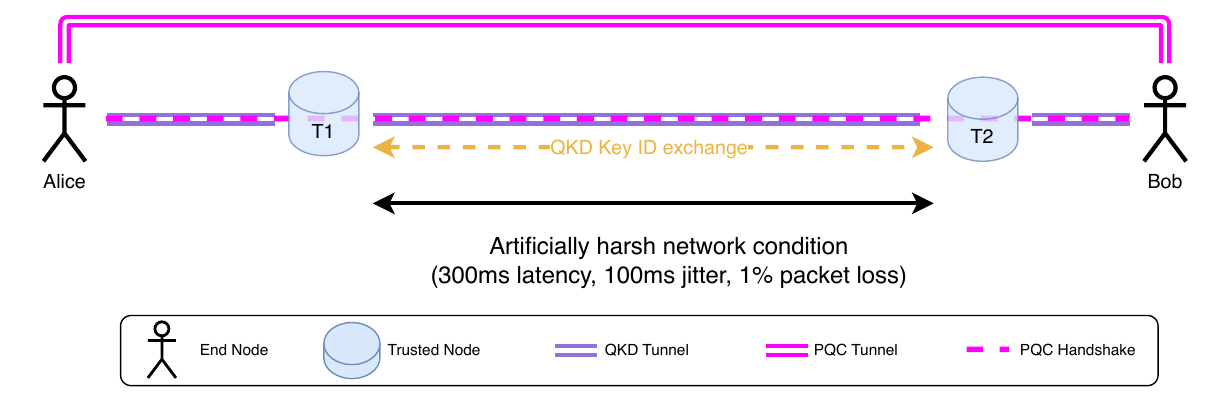}
    \caption{Simulated testbed with QKD daisy chain and challenging network conditions}
    \label{fig:testbed-harsh}
\end{figure}

To evaluate behavior under degraded connectivity, two end nodes were connected through a daisy chain of two intermediate trusted nodes. The link between the trusted nodes was configured with 300 ms latency, 100 ms jitter, and 1\% packet loss. These impairments affected both the QKD key negotiation performed by Arnika and the PQC handshake executed through the tunnel.

The simulation exposed race conditions in Arnika and Rosenpass that produced failures to establish a common key. For this study, the open-source implementations were partly modified and the issues were reported to the upstream projects\footnote{References withheld for review}, enabling correct operation under degraded connectivity.

Across 100 runs, the mean time to provision a data tunnel secured by one QKD paths over a degraded connection was 11.6s.

\subsubsection{Test 5 - Simulated QKD malfunction}

To verify the fail-safe mechanism a malfunction of a single QKD node was simulated. The topology from test 2 was deployed and after the data link became available, the first QKD container was terminated. This would not immediately cause a disruption of the data tunnel, instead Arnika failed to negotaiate a QKD key and disrupted the WireGuard tunnel between \textit{end node A} and \textit{trusted node 01} after 180 seconds. Another 60s later (at 240s), the WireGuard hop would try to create a new session key, however fail due to the randomly injected PSK from Arnika. At 300s the grace period of the WireGuard tunnel is passed and no further PQC handshakes could be exchanged. Rosenpass would try a new (but failing) exchange at 360s and inject random keys at 420s. At 480s the WireGuard data tunnel would start a new session handshake but fail, disrupting the data tunnel after 540s.

Across 100 runs, the mean time between terminating the QKD simulator and the disruption of the data tunnel was 548.42s.

\section{Security Evaluation}
\label{sec:security-evaluation}

This section presents a security evaluation of the system, emphasizing forward secrecy and resistance to "harvest-now, decrypt-later" attacks. Essentially, compromise of the end-to-end WireGuard tunnel would require an adversary to simultaneously
\begin{enumerate}
    \item break classical cryptographic mechanisms,
    \item compromise trusted nodes or break QKD devices, and
    \item break the PQC primitives.
\end{enumerate}
The resulting layered structure establishes defense-in-depth, raising the effective attack complexity.

\textbf{(1) Classical Cryptography}\label{sec:classical}

The system inherits the formally verified security guarantees of WireGuard. Prior analyses confirm that its Noise-based key exchange achieves end-to-end authenticated key exchange with perfect forward secrecy~\cite{donenfeld2017wireguard,DBLP:conf/eurosp/KobeissiNB19,lipp2019mechanised,dowling2018cryptographic}. The optional pre-shared key (PSK) mode further strengthens confidentiality by introducing an additional secret input that must be compromised alongside the session handshake to break confidentiality. Injecting a PSK does not degrade security; an adversary would need to compromise both WireGuard's session handshake protocol and the PSK material to recover session keys and break confidentiality of the transported data~\cite{dowling2018cryptographic,cremers2012beyond}.

\textbf{(2) Quantum Key Distribution (QKD)}\label{sec:qkd}

The system periodically refreshes the PSK using keys obtained from QKD devices that conform to the ETSI GS QKD 014 interface standard~\cite{ETSI014}, which ensures standardized, authenticated key delivery between QKD modules and key management entities, preventing substitution or injection of unauthorized key material. The evaluation scope is limited to protocol-level guarantees; hardware-specific vulnerabilities such as detector blinding or side-channel leakage are excluded. Authentication between two QKD endpoints is established during the initial deployment with vendor-specific methods. Any man-in-the-middle attacks on the quantum layer would stop the QKD devices from producing new key material. Hence, if a key can be obtained via the ETSI GS QKD 014 interface, this specific key can be assumed to be authentic and uncompromised for this specific QKD link.

Communication between \textit{Arnika} instances is neither encrypted nor authenticated. This is not considered a security weakness, as only ETSI 014 key IDs are exchanged. An adversary could at most disrupt operation, but would not gain access to any key material at any point.

\textbf{(3) Post-Quantum Cryptography (PQC)}\label{sec:pqc}

The PQC components rely on established security proofs and NIST-endorsed cryptographic standards~\cite{NISTPQC,rosenpass_whitepaper}. Rosenpass integrates Classic McEliece-based key encapsulation for authentication~\cite{bernstein2008attacking} and CRYSTALS-Kyber for key exchanges. The resulting shared key enjoys post-quantum end-to-end security and is authentication.

Crucially, the PQC key exchange itself benefits from protection via the QKD infrastructure. Sending the PQC key exchange via the hop-to-hop VPN links secured via WireGuard using QKD keys as PSK, effectively mitigates  threats posed by passive attackers who might attempt to store encrypted traffic in anticipation of future vulnerabilities in classical or PQC cryptographic methods (\textit{harvest now, decrypt later} attacks). Furthermore, for active adversaries to interfere with the PQC key exchange, the adversary effectively needs to compromise the QKD-WireGuard links, the trusted nodes, or the end nodes. Given the security of WireGuard against active adversaries, unless the adversary is capable of breaking both the classical key exchange of WireGuard and the QKD protocol, compromise of the trusted node remains the only viable method for an attack. For an adversary having access to a trusted node, the security of the WireGuard layer for the data exchange, is secured by the end-to-end security properties of the classical key exchange of WireGuard and of Rosenpass. Hence, an adversary would again be capable to break both classical and post-quantum secure protocols.
Consequently, successful decryption would necessitate actively compromising the QKD system, and breaking both classical and PQC key exchanges and thus our layer approach substantially increases the security against potential adversaries.

\section{Discussion}

The prototype validation demonstrates that multi-hop QKD-hardened tunnels can be constructed with minimal overhead and without stressing cryptographic components. End-to-end setup time remains dominated by transport-layer behavior rather than QKD or PQC processing, indicating that the architecture does not introduce bottlenecks beyond those inherent to the underlying tunnel mechanism. The low control-plane traffic confirms that the design imposes negligible overhead during key establishment, even when traversing several trusted nodes.

The scaling simulation shows that path length exerts limited influence on initialization latency. Because QKD tunnels are established in parallel, the critical factor is the slowest hop rather than the aggregate number of hops. Once QKD secured tunnels are available, the PQC handshake traverses the chain with a fixed, small payload, yielding setup times that remain stable even at 100 trusted nodes. This indicates that long terrestrial QKD paths do not impose compounding delays on the layered construction.

The dual-link simulation confirms that independent QKD daisy chains allow PSKs to be combined at the endpoints, raising the compromise threshold. The design provides cryptographic agility by allowing multiple PQC implementations to run in parallel at the endpoints, enabling incremental migration without requiring simultaneous upgrades of the entire network.

The degraded-connectivity test shows that the architecture remains functional even when link quality deteriorates. Although latency, jitter, and loss slow initial tunnel establishment, the system maintains correctness once implementation-level race conditions are addressed. This suggests that robustness under adverse conditions depends primarily on software behavior, not architectural constraints.

The final experiment confirms controlled failure propagation when QKD becomes unavailable. Each layer fails according to its own timeout and re-key schedule. This behavior mitigates downgrade attacks by preventing extended usage of the same key material while simultaneously providing a bounded window in which remediation is possible.

Overall, our simulations indicate that the architecture scales, tolerates cryptographic and network heterogeneity, and degrades in a predictable, security-preserving manner.

\section{Conclusion}
\label{sec:conclusion}

This work presented a modular architecture that integrates classical, post-quantum, and quantum cryptographic mechanisms without reliance on Key Management Systems with forwarding capabilities. By combining QKD-derived pre-shared keys for link-level protection with PQC-enhanced WireGuard tunnels via Rosenpass, the design achieves layered forward secrecy and end-to-end confidentiality across multi-hop networks containing intermediate nodes.

The integration of QKD strengthens PQC’s mathematical resilience against \textit{harvest-now, decrypt-later} attacks while mitigating its intrinsic limitations in authenticity, distance, and scalability. Security evaluation shows that independent cryptographic layers maintain system integrity and prevent cascading failures, even if one layer is compromised.

Experimental validation across multiple simulations confirmed the system’s efficiency and stability. The architecture exhibits minimal computational and memory overhead, establishes tunnels in under 15 seconds, and maintains stable connectivity under degraded connectivity.

Collectively, these findings establish a practical path toward deployable, quantum-safe network infrastructures. The design’s reliance on open-source components and compliance with standardized interfaces allows transparent integration with existing QKD ecosystems and facilitates gradual migration toward fully quantum-resilient communication systems.

\subsection{Future Work}

Future work will expand the architecture to carrier-scale simulations and field deployments as quantum communication infrastructure matures. Large field trials across operational QKD networks will test path diversity, routing dynamics, and long-term stability under production traffic patterns. Additionally, alternative inter-hop link technologies should be evaluated, including LTE-based links and long-distance nodes where QKD material is generated via satellite.

Security and interoperability will advance through the addition of authentication, authorization, and accounting mechanisms (AAA), along with systematic evaluation of interactions with carrier-grade routing and firewall systems.

Resource usage requires deeper analysis and system level evaluations. The modular architecture permits substitution studies, including MACsec or IPsec in place of WireGuard and alternative PQC handshakes based on different cryptographic primitives. PQC software may also be extended to expose an ETSI-014 interface, allowing integration with existing infrastructure.

Sustained scrutiny of cryptographic components remains necessary. Work will include a security assessment of Arnika and periodic re-evaluation of emerging PQC standards and deployments, including PQC-TLS and outcomes of forthcoming PQC competitions.

\section*{Acknowledgment}

We thank Stephan Laschet, Florian Kutschera (AIT), the Rosenpass~e.V. team, and Benjamin Lipp for their support and comments. This work was supported by Q-net-Q, funded by the European Union's Digital Europe Programme (DEP) under grant agreement No 101091732 and co-funded by the German Federal Ministry of Education and Research (BMBF), QCI-CAT funded by DEP under grant aggreement No 101091642 and the National Foundation for Research, Technology and Development. This work has received funding from the European Union’s Horizon Europe research and innovation program under No. 101114043 ("QSNP"). The views expressed are those of the authors and do not necessarily reflect the views of the funding agencies.

\bibliographystyle{alpha}
\bibliography{base,dblp}

\end{document}